\documentclass[1p,times,authoryear]{article}        
\usepackage{graphicx}

 \usepackage{mathptmx}   
\usepackage{amsmath}
\usepackage{amssymb}
\usepackage{latexsym}
\usepackage{epsfig}
\usepackage{amsfonts}
\usepackage[latin1]{inputenc}
\usepackage{float}
\usepackage{mathrsfs}
\usepackage{rotating}
\usepackage{natbib}

\newcommand{\ci}{\perp\!\!\!\perp}



\begin{document}

\title{Targeted smoothing parameter selection for estimating average causal effects}
\author{\baselineskip 19pt Jenny H\"aggstr\"om\footnote{Corresponding address: Department of Statistics, Ume{\aa} School of Business and Economics, Ume{\aa} University, SE-90187 Ume{\aa}, Sweden. E-mail: jenny.haggstrom@stat.umu.se  Tel: +46 90 7869318. Fax: +46 90 7866614.}\ \  and Xavier de Luna  \\
\\
Department of Statistics, Ume{\aa} School of Business and Economics, \\ Ume{\aa} University, SE-90187 Ume{\aa}, Sweden
}

\date{}

\maketitle

\begin{abstract}
The non-parametric estimation of average causal effects in observational studies often relies on controlling for confounding covariates through smoothing regression methods such as kernel, splines or local polynomial regression. Such regression methods are tuned via smoothing parameters which regulates the amount of degrees of freedom used in the fit. In this paper we propose data-driven methods for selecting smoothing parameters when the targeted parameter is an average causal effect. For this purpose, we propose to estimate the exact expression of the mean squared error of the estimators. Asymptotic approximations indicate that the smoothing parameters minimizing this mean squared error converges to zero faster than the optimal smoothing parameter for the estimation of the regression functions.
In a simulation study we show that the proposed data-driven methods for selecting the smoothing parameters yield lower empirical mean squared error than other methods available such as, e.g., cross-validation.
\end{abstract}

\section{Introduction}
In observational studies where the interest lies in estimating the average causal effect of a binary treatment $z$ on an outcome of interest $y$, non-parametric estimators are typically based on controlling for confounding covariates $x$ with smoothing regression methods (nearest neighbour, kernel, splines, local polynomial regression, series estimators; see, e.g., the review by \citeauthor*{imbwo09}, \citeyear{imbwo09}). A useful modeling framework in this context was introduced by \citet{neyman23} and \citet{rubin74}, where in particular two potential outcomes are considered for each unit in the study, the outcome that would be observed if the unit is treated, $y(1)$, and the outcome that would be observed if the unit is not treated, $y(0)$. The causal effect at the unit level is defined as $y(1)-y(0)$. Population parameters are targeted by the inference, and we focus here on average causal effects of the type $E(y(1)-y(0))$, where the expectation is taken over a given population of interest. Inference on such expectations is complicated by the fact that the two potential outcomes are not observed for all units in the sample (missing data problem) and assumptions, e.g., on the missingness mechanism must be made in order for the parameter of interest to be identified. In this paper, we consider situations described in Section 2, where the causal effect conditional on an observed covariate $x$ (or a score function summarizing a set of observed covariates), $E(y(1)\mid x)-E(y(0)\mid x)$, is identified and can be estimated by fitting two curves, functions of $x$, $E(y(1)\mid x,z=1)$ and $E(y(0)\mid x,z=0)$ non-parametrically. An estimate of the targeted average causal effect is obtained by averaging the estimated curves over the relevant distribution for $x$ to target $E(y(1)-y(0))=E(E(y(1)\mid x))-E(E(y(0)\mid x))$, where the missing outcomes are imputed by predictions from the fitted curves. A tuning parameter for each fitted curve is used to regulate the smoothness of the fit. \cite{cheng94} showed that
when using kernel regression to estimate the average of a curve, say here $E(E(y(1)\mid x))$, with missing $y(1)$ for some units, as described above, then the optimal (in mean squared error, MSE, sense) smoothing parameter for the estimation of the regression curve $E(y(1)\mid x,z=1)$ is not optimal for the estimation of the average $E(E(y(1)\mid x))$.
More precisely the optimal rate of convergence towards zero of the smoothing parameter (when the sample size increases) is different in both situations, and one need typically to asymptotically undersmooth $E(y(1)\mid x,z=1)$ when targeting $E(E(y(1)\mid x))$. We show in this paper that a similar result holds when using local linear regression instead of kernel regression, and when two curves (implying the choice of two tunining parameters), are fitted and then averaged to target $E(y(1)-y(0))$.

As a main contribution of the paper, we propose a novel data-driven method geared for selecting the smoothing parameters which minimizes the mean squared error of non-parametric estimators of the average causal effect. \cite*{inr05} also proposes a data-driven method based on the estimation of this mean squared error. The two estimators are, however, different. While \cite{inr05} estimates an asymptotic approximation of the population MSE which involves the estimation of the propensity score, the probability of ending up in one of the treatment groups (say $z=1$) given the covariates, our estimator targets the exact population MSE by using a double smoothing technique previously used by \cite*{hardleetal92} for estimating regression curves and \cite{hagg10} in semi-parametric additive models. Note that \cite{frolich05} also derived asymptotic approximation of MSE to obtain smoothing parameter selectors although those were outperformed by cross-validation in finite sample simulations. With simulations we study the finite sample properties of the different data-driven methods. The results suggest that the cross-validation choice, which is known to be optimal in MSE sense to estimate smooth curves \citep{fan92}, can indeed be improved by using either \cite{inr05} or our proposal, with the latter often being superior. 

In the next section we introduce the potential outcome framework dating back to \citet{neyman23} and \citet{rubin74}, which allows us to define the parameter of interest, the average causal effect, and commonly used identifying assumptions and estimators. The selection of smoothing parameters is discussed in Section 2, where we present asymptotic results based on the use of local linear regression. We also introduce in this section a novel data-driven method. Section 3 presents a simulation study. The paper is concluded in Section 4.

\section{Model and estimation}
\subsection{Neyman-Rubin model for causal inference}
Suppose we have $n$ units $i$ in a study, a random sample from a population of interest for which we observe a binary treatment assignment $z_i$, a real valued outcome $y_i$ and a set of covariates $\mathbf{x}_i$. Thus,
$$
z_i=\left\{ \begin{array}{ll}
1 & \textrm{if unit $i$ recieves treatment 1},\\
0 & \textrm{if unit $i$ recieves treatment 0}\\
&  \, \textrm{(possibly no treatment, control group)}.
\end{array} \right.
$$
\noindent
The causal effect of treatment $z_i=1$ versus treatment $z_i=0$ on the response variable $y$ for unit $i$ is defined as 
$
\tau_i=y_i(1)-y_i(0),
$
\noindent
with $y_i(1)$ and $y_i(0)$ the potential outcomes for unit $i$, i.e. $y_i(1)$ is the response that would be observed for unit $i$ if given treatment $z_i=1$ and $y_i(0)$ the response if given treatment $z_i=0$. The observed response for unit $i$ is then
$
y_i=y_i(0)(1-z_i)+y_i(1)z_i.
$
The individual causal effect $\tau_i$ is not observable since unit $i$ can only receive one of the two treatments. Typically, 
the parameter of interest is a population average causal effect,
\begin{equation}
\tau=E\big(y_i(1)-y_i(0)\big).\notag
\end{equation}
\noindent
If treatment assignment is not randomized, 
$\tau$ is identified if we have available a set of covariates $\mathbf{x}_{i}=(x_{i1},\ldots, x_{id})^{T}$ not affected by treatment assignment and such that the following assumptions hold,
$$
y_i(1), y_i(0)  \ci z_i|\mathbf{x}_i, 
$$
often called unconfoundedness assumption,
and
$$
0<\Pr(z_i=1|\mathbf{x}_i)<1,
$$
often called overlap assumption.
We have unconfoundedness if all covariates affecting both treatment assignment and the potential outcomes are included in $\mathbf{x}_i$. The assumption of overlap states that, for a unit with covariate vector $\mathbf{x}_i$, the probability  of receiving either treatment should be bounded away from 0. Under these assumptions identifiability of $\tau$ is then a consequence of 
\begin{align}\label{davidstar}
\tau&=E\big(y_i(1)-y_i(0)\big)\notag\\
&=E\big(E(y_i(1)|\mathbf{x}_i)-E(y_i(0)|\mathbf{x}_i)\big) \nonumber \\ 
&=E\big(E(y_i(1)|z_i=1,\mathbf{x}_i)-E(y_i(0)|z_i=0,\mathbf{x}_i)\big) \nonumber \\ 
&=E\big(E(y_i|z_i=1,\mathbf{x}_i)-E(y_i|z_i=0,\mathbf{x}_i)\big).
\end{align}

\noindent
In the sequel we focus on the case $d=1$ since when $d>1$, the covariate vector $\mathbf{x}_i$ can be replaced by a scalar, e.g., $p(\mathbf{x}_i)=\Pr(z_i=1|\mathbf{x}_i)$, the propensity score (\citeauthor{rosrub83}, \citeyear{rosrub83}, \citeauthor{hansen08}, \citeyear{hansen08}). Indeed, 
 \citet{rosrub83} showed that it is sufficient to condition on the propensity score, i.e. under the above assumptions we have
$
y_i(1), y_i(0)  \ci z_i|p(\mathbf{x}_{i}),
$
and
$
0<\Pr(z_{i}=1|p(\mathbf{x}_{i}))<1.
$
In applications the propensity score need to be modelled and fitted to the data. Typically parametric models are used, although these do not need to be correctly specified as shown in \cite{IW:10}.

\subsection{Estimating average causal effects}
Let $\beta_0(x_i)=E(y_i|z_i=0,{x}_i)$ and $\beta_1(x_i)=E(y_i|z_i=1,{x}_i)$ be unknown smooth functions, $Var(y_i|x_i, z_i)=\sigma_{\epsilon}^2$. Note that the assumption of constant conditional variance could be relaxed without changing in essence the results of this paper. We consider this assumption to alleviate the notational burden. From (\ref{davidstar}), we have that
$$
\tau=E\big(\beta_1(x_i)\big)-E\big(\beta_0(x_i)\big).
$$
\noindent
Thus, a natural way to estimate $\tau$ is to first estimate the two regression functions $\beta_1(x_i)$ and $\beta_0(x_i)$, based on the treated and the non-treated, respectively, and then take the average over all the observed $x_i$s of the differences between the estimated functions. This estimator of $\tau$ is called the imputation estimator in \cite{inr05}. They use series estimators for estimating the regression functions but any smoother, e.g.  nearest neighbour, kernel, splines and local
polynomial regression \citep[p.~14--45]{fangijbels96}, may be used. 

Denote $\mathbf{y}^0=(y_{1}^0,\ldots,y_{n_0}^0)^T$ and $\mathbf{x}^0=(x_{1}^0,\ldots,x_{n_0}^0)^T$ the observed response and covariate for the $n_0$ units with treatment $z_i=0$, and similarly $\mathbf{y}^1=(y_{1}^1,\ldots,y_{n_1}^1)^T$ and $\mathbf{x}^1=(x_{1}^1,\ldots,x_{n_1}^1)^T$ for the $n_1$ units with treamtment $z_i=1$.
The smoothers cited above are linear in the sense that
the corresponding estimator of $\beta_j(\mathbf{x})=(\beta_j(x_1),\ldots,\beta_j(x_n))^T$, can be written as 
\begin{align*}
\hat{\beta}_{j}^{h_j}(\mathbf{x})&=S_{j}^{h_j}[\mathbf{x}]\mathbf{y}^{j}, \ \ j=0,1,  \\
\end{align*}
where $\mathbf{x}=(\mathbf{x}^{0T},\mathbf{x}^{1T})^T$ and $S_{j}^{h_j}[\mathbf{x}]$ the smoothing matrix regressing $\mathbf{y}^j$ on $\mathbf{x}^j$, using smoothing parameter $h_j$. The imputation estimator of $\tau$ mentioned above is
\begin{equation}
\hat{\tau}^{imp}=\frac{1}{n}\sum_{i=1}^n\hat{\tau}^{imp}(x_i)=\frac{1}{n}\sum_{i=1}^n\big(\hat{\beta}_{1}^{h_1}(x_i)-\hat{\beta}_{0}^{h_{0}}(x_i)\big).\notag
\end{equation}
\noindent
In this paper we base our results on a specific linear smoother, the local linear regression smoother, although we anticipate that most results should hold for any other linear smoother.

Local linear regression \citep{cleveland79,fangijbels96}, consists in fitting a straight line at every $x_{i}$, $i=1,\ldots,n$, using only the part of data that is deemed to be sufficiently close to the target point $x_{i}$. 
Consider estimating the regression function $\beta_j(\cdot)$, $j=1,0$. The fit, at $x_i$, is 
$$
\hat{\beta}_j^{h_j}(x_i)=\mathbf{e}_{1}^T(\mathbf{X}_{i}^{jT}\mathbf{W}_i^{h_j}\mathbf{X}_{i}^{j})^{-1}\mathbf{X}_{i}^{jT}\mathbf{W}_i^{h_j}\mathbf{y}^j=S_j^{h_j}[x_i]\mathbf{y}^j
$$
where $\mathbf{e}_{1}=(1,0)^T$, 
$$
\mathbf{X}_{i}^{j}=
\left( \begin{array}{cc}
1 & (x_{1}^j-x_i) \\
\vdots & \vdots \\
1 & (x_{n_j}^j-x_i) 
\end{array} \right )
$$
and
$$
\mathbf{W}_i^{h_j}=\mbox{diag}(K\big((x_{1}^j-x_i)/b_{ji}\big)/b_{ji}, \ldots, K\big((x_{n_j}^j-x_i)/b_{ji}\big)/b_{ji}).
$$
$K(\cdot)$ is a kernel function such that $\int K(u) du=1$ and $\int u K(u)du=0 $. An example is the tricube kernel defined as 
$$
K(u)=\left \{ \begin{array}{ll}
\frac{70}{81}(1- |u|^{3})^{3}, & \textrm{if $|u|<$ 1}\\
~0, & \textrm{if $|u|\geq$ 1}
\end{array} \right \}.
$$
The definition of $b_{ji}$, $i=1,\ldots,n$, depends on the type of bandwidth we use. With a constant bandwidth  $b_{j1}=\cdots=b_{jn}=h_j$. For a nearest neighbor type bandwidth, assuming no ties, $b_{ji}$ is the Euclidian distance from $x_i$ to the $(h_jn_j)$:th nearest among the $x_{k}^j$:s for $x_{k}^j\neq x_i,\, \, h_j\in [1/n_j,1]\,, k=1,\ldots,n_j$, and the smoothing parameter $h_j$ is the proportion of observations being used to produce the local fit.

\section{Selection of smoothing parameters}

\subsection{Mean squared errors}
Many smoothing parameter selection methods are developed with the purpose of estimating the regression function $\beta_j(x_i)$, $j=1,0$, and attempts to select the smoothing parameter minimizing the average conditional mean squared error:
\begin{align}\label{msebeta}
&\frac{1}{n_j}\sum_{i=1}^{n_j}Var\big(\hat{\beta}_j^{h_j}(x_{i}^j)|\mathbf{x}^j\big)+\frac{1}{n_j}\sum_{i=1}^{n_j}E\big(\hat{\beta}_j^{h_j}(x_{i}^j)-\beta_j(x_i^j)|\mathbf{x}^j\big)^2\notag\\
=&\frac{\sigma_{\epsilon}^2}{n_j}\sum_{i=1}^{n_j}S_{j}^{h_j}[x_{i}^j]S_{j}^{h_j}[x_{i}^j]^T+\frac{1}{n_j}\sum_{i=1}^{n_j}\bigg(S_{j}^{h_j}[x_{i}^j]\beta_j(\mathbf{x}^j)-\beta_j(x_{i}^j)\bigg)^2.\notag\\
\end{align}
\noindent
One frequently used selection procedure that attempts to select the smoothing parameter minimizing (\ref{msebeta}) is leave-one-out cross-validation.  In this setting, cross-validation selects the smoothing parameter $h_j$ minimizing
\begin{equation}\label{cv}
\frac{1}{n_j}\sum_{i=1}^{n_j}\big(y_{i}^j-\hat{\beta}_j^{h_j,-i}(x_{i}^j)\big)^2,
\end{equation}
\noindent
where $\hat{\beta}_j^{h_j,-i}(x_{i}^j)$ is the cross-validated estimate at $x_{i}^j$ computed without $(x_{i}^j,y_{i}^j)$. Asymptotically, for local linear regression, the smoothing parameter minimizing (\ref{msebeta}) is proportional to $n_j^{-1/5}$ \citep{fan92}, and, hence, proportional to $n^{-1/5}$ since $n_j=n\Pr(z=j)+o_p(n)$. However, it is known that for estimating a functional of  $\beta_j(x_i)$ such as $E(\beta_j(x_i))$, the smoothing parameter minimizing (\ref{msebeta}) is not optimal, in the sense that it does not result in $\sqrt{n}$-consistent estimation of the functional \citep[e.g.,][]{cheng94}. \cite{inr05} suggest that one should select $h_0$ and $h_1$ by minimizing the conditional mean squared error of $\frac{1}{n}\sum_{i=1}^n\hat{\beta}_j^{h_j}(x_i)$, for $j=0,1$ respectively, i.e.
\begin{align}\label{mseinr}
MSE_{\bar{\hat{\beta}}_j}=&\frac{\sigma_{\epsilon}^2}{n^2}\sum_{i=1}^n\sum_{k=1}^nS_{j}^{h_j}[x_i]S_{j}^{h_j}[x_k]^{T}\notag\\
&+\frac{1}{n^2}\bigg[\sum_{i=1}^n\bigg(S_{j}^{h_j}[x_i]\beta_j(\mathbf{x}^j)-\beta_j(x_i)\bigg)\bigg]^2.
\end{align}

\noindent
We argue that, in order to estimate $\tau$ optimally, it may be more suitable to select the combination of ($h_1,h_0$) minimizing the conditional mean squared error of $\hat{\tau}^{imp}$

\begin{align}\label{msetauhat}
MSE_{\hat{\tau}}=&\frac{\sigma_{\epsilon}^2}{n^2}\sum_{i=1}^{n}\sum_{j=1}^n\bigg(S_1^{h_1}[x_i]S_{1}^{h_1}[x_j]^T\notag\\
&\,\,\,\,\,\,\,\,\,\,\,\,\,\,\,\,\,\,\,\,\,\,\,\,\,+S_{{0}}^{h_{0}}[x_i]S_{0}^{h_{0}}[x_j]^T\bigg)\notag\\
&+\bigg[\frac{1}{n}\sum_{i=1}^n\bigg(\big(S_{1}^{h_1}[x_i]\beta_1(\mathbf{x}^1)-\beta_1(x_i)\big)\notag\\
&\,\,\,\,\,\,\,\,\,\,\,\,\,\,\,\,\,\,\,\,\,\,\,\,\,-\big(S_{0}^{h_0}[x_i]\beta_0(\mathbf{x}^0)-\beta_0(x_i)\big)\bigg)\bigg]^2.\notag\\
\end{align}
\noindent
Note that 

\begin{align*}
MSE_{\hat{\tau}}=&MSE_{\bar{\hat{\beta}}_1}+MSE_{\bar{\hat{\beta}}_0}\\
&-2\bigg(\frac{1}{n}\sum_{i=1}^n\big(S_{1}^{h_1}[x_i]\beta_1(\mathbf{x}^1)-\beta_1(x_i)\big)\bigg)\\
&\times\bigg(\frac{1}{n}\sum_{i=1}^n\big(S_{0}^{h_0}[x_i]\beta_0(\mathbf{x}^0)-\beta_0(x_i)\big)\bigg).\\
\end{align*}
Hence, criterion (\ref{msetauhat}) differs from (\ref{mseinr}) when both average bias terms in the latter expression are different from zero.

\subsection{Asymptotics}
Asymptotic approximations can be used to describe optimal bandwidth choices as the sample size tends to infinity.
The results presented here are deduced in Appendix B, where regularity conditions also used in \cite{ruwa94} are given. 
For local linear regression with constant bandwidth such that $h_j\rightarrow 0$ and $nh_j \rightarrow \infty$ as $n\rightarrow \infty$ we have the following approximations for
the conditional bias and variance of $\frac{1}{n}\sum_{i=1}^n\hat{\beta}_j^{h_j}(x_i)$. For $j=1,0$, 
\begin{align}\label{condbias.eq}
E&\bigg(\frac{1}{n}\sum_{i=1}^n\hat{\beta}_j^{h_j}(x_i)-\frac{1}{n}\sum_{i=1}^n\beta_j(x_i)|\mathbf{x}\bigg)\notag\\
&=B_1(j)h_j^2+o_p(h_j^2),
\end{align}
and
\begin{align}\label{condvar.eq}
Var\bigg(\frac{1}{n}\sum_{i=1}^n\hat{\beta}_j^{h_j}(x_i)|\mathbf{x}\bigg)=&\frac{V_1(j)}{n}+\frac{V_2(j)}{n^2h_j}+V_3(j)\frac{h_j^2}{n}\notag\\
&+o_p\big(n^{-1}+n^{-2}h_j^{-1}+n^{-1}h_j^2\big),
\end{align}
with constants

\begin{align*}
B_1(j)=&\frac{1}{2}\int \beta_j^{(2)}(x)f(x)dx\int u^2K(u)du,\\
V_1(j)=&\sigma_{\epsilon}^2\int \frac{f(x)}{Pr(z=j|x)}dx,\\
V_2(j)=&\sigma_{\epsilon}^2\int K(u)^2du\int\frac{1}{Pr(z=j|x)}dx,\\ 
V_3(j)=&-2\sigma_{\epsilon}^2\int u^2K(u)du\int \frac{f^{(1)}(x)^2}{f(x)Pr(z=j|x)}dx\\
&-2\sigma_{\epsilon}^2\int u^2K(u)du\int\frac{f^{(1)}(x)P^{(1)}(z=j|x)}{Pr(z=j|x)^2}dx,\\
\end{align*}

\noindent
where $\beta_j^{(m)}(x)$ the $m$:th derivative of the function $\beta_j(x)$ and $f(x)$ is the density of $x$.
Hence,
\begin{align}\label{mseinrasym}
MSE_{\bar{\hat{\beta}}_j}=&\frac{V_1(j)}{n}+\frac{V_2(j)}{n^2h_j}+V_3(j)\frac{h_j^2}{n}+B_1^2(j)h_j^4\notag\\
&+o_p\big(n^{-1}+n^{-2}h_j^{-1}+n^{-1}h_j^2+h_j^4\big)
\end{align}
\noindent 
and
\begin{align}\label{msetauhatasym}
MSE_{\hat{\tau}}=&\frac{V_1(1)+V_1(0)}{n}+\frac{V_2(1)}{n^2h_1}+\frac{V_2(0)}{n^2h_0}\notag\\
&+V_3(1)\frac{h_1^2}{n}+V_3(0)\frac{h_0^2}{n}+B_1^2(1)h_1^4\notag\\
&+B_1^2(0)h_0^4-2B_1(1)B_1(0)h_1^2h_0^2\notag\\
&+o_p\big(n^{-1}+n^{-2}h_1^{-1}+n^{-2}h_0^{-1}+n^{-1}h_1^2\notag\\
&\,\,\,\,\,\,\,\,\,\,\,\,\,\,\,\,\,\,\,\,\,+h_0^2n^{-1}+h_1^4+h_0^4+h_1^2h_0^2\big).\\\notag
\end{align}

\noindent

Let us first consider
the optimal smoothing parameter for estimating $E(\beta_j(x))$ and assume $nh_j^3\rightarrow 0$ as $n\rightarrow \infty$, $j=0,1$. An asymptotic approximation to the bandwidth minimizing (\ref{mseinrasym}) is 
$$
h_j^{opt}=\text{arg}\min_{h_j}\frac{V_2(j)}{n^2h_j}+B_1^2(j)h_j^4=\bigg(\frac{V_2(j)}{4B_1^2(j)}\bigg)^{1/5}n^{-2/5}.
$$
Hence, the optimal rate of convergence is here faster than $n^{-1/5}$, the optimal rate for the estimation of the regression function $\beta_j(\cdot)$. A similar result was shown in \cite{cheng94}
for kernel regression. Turning to the minimization of (\ref{msetauhatasym}), this must be done simultaneously in $h_0$ and $h_1$. A reasonable assumption, however, is that these two smoothing parameters have same rate of convergence to zero. Under this assumption we may replace $h_1$ by $ch_0$, for $c$ a constant, in (\ref{msetauhatasym}). Minimizing the latter for $h_0$ yields as above an optimal rate of convergence for $h_0$ (and hence $h_1$) of $n^{-2/5}$.

Another related result, deduced from (\ref{condbias.eq}) and (\ref{condvar.eq}), is that as $n \rightarrow \infty$, if $h_j\propto n^r$, for $-1< r < -1/4$, then (see Appendix B)
\begin{eqnarray}
E\left[\sqrt{n}(\bar{\hat{\beta}}_j-E(\beta_j(x_i)))\mid \mathbf{x}\right]&=&o_p(1),  \label{asymnorminra}\\
E\left[\sqrt{n}(\hat{\tau}^{imp}-\tau)\mid \mathbf{x}\right]&=&o_p(1),\label{asymnorminrb}\\
Var\left[\sqrt{n}(\bar{\hat{\beta}}_j-E(\beta_j(x_i)))\mid \mathbf{x}\right]&=&V_1(j)+o_p(1),\label{asymnorminrc}\\
Var\left[\sqrt{n}(\hat{\tau}^{imp}-\tau)\mid \mathbf{x}\right]&=&V_1(0)+V_1(1)\notag\\
&&\,+o_p(1).\label{asymnorminrd}
\end{eqnarray}
\noindent
The results above show that selecting the smoothing parameters minimizing (\ref{mseinr}) will lead to $\sqrt{n}$-consistent estimation of $\tau$. This is in accordance with previous results (e.g., \citeauthor{speckman88}, \citeyear{speckman88}) where it was shown that asymptotic undersmoothing of the regression function is needed for the $\sqrt n-$consistent estimation of a functional of the regression function.

\subsection{Estimating MSEs}
\cite{inr05} propose the following estimator of (\ref{mseinr})

\begin{align}\label{msehatinr}
\widehat{MSE}_{\bar{\hat{\beta}}_j}^{INR}&=\frac{\hat{\sigma_{\epsilon}}^2}{n^2}\sum_{i=1}^n\sum_{k=1}^nS_{j}^{h_j}[x_i]S_{j}^{h_j}[x_k]^{T}\notag\\
&+\frac{1}{n^2}\bigg[\sum_{i=1}^{n_j}\frac{1}{\hat{p}(x_{i}^j)}\bigg(y_{i}^j-\hat{\beta}_j^{h_j}(x_{i}^j)\bigg)\bigg]^2\notag \\
&-\frac{\hat{\sigma_{\epsilon}}^2}{n^2}\hat{{\mathbf p}}_j^T\bigg(I_{n_j}-S_{j}^{h_j}[\mathbf{x}^j]\bigg)\bigg(I_{n_j}-S_{j}^{h_j}[\mathbf{x}^j]\bigg)^T\hat{{\mathbf p}}_j, \\\notag
\end{align}

\noindent
where $\hat{{\mathbf p}}_j=(1/\hat{p}(x_{1}^j),\ldots,1/\hat{p}(x_{n_j}^j))^T$ and $I_{n_j}$ is the $n_j\times n_j$ identity matrix. It is worth noting that one need to estimate the propensity score \citep{IW:10}, in addition to $\sigma_{\epsilon}^2$, in order to use this selection procedure. The error variance $\sigma_{\epsilon}^2$ may be estimated by

$$
\hat{\sigma_{\epsilon}}^2=\frac{(\mathbf{y}^j-S_j^{h_{\epsilon}}[\mathbf{x}^j]\mathbf{y}^j)^T(\mathbf{y}^j-S_j^{h_{\epsilon}}[\mathbf{x}^j]\mathbf{y}^j)}{n-\text{trace}(2S_j^{h_{\epsilon}}[\mathbf{x}^j]-S_j^{h_{\epsilon}}[\mathbf{x}^j]S_j^{h_{\epsilon}}[\mathbf{x}^j])},
$$
where $h_{\epsilon}$ could be equal to $h_j$ or selected separately, see e.g. \cite{opetal95} for further discussion on this issue.

We propose below novel double smoothing estimators of (\ref{mseinr}) and (\ref{msetauhat}), respectively:

\begin{align}\label{msehatinrds}
\widehat{MSE}_{\bar{\hat{\beta}}_j}^{DS}=&\frac{\hat{\sigma_{\epsilon}}^2}{n^2}\sum_{i=1}^n\sum_{k=1}^nS_{j}^{h_j}[x_i]S_{j}^{h_j}[x_k]^{T}\notag\\
&+\frac{1}{n^2}\bigg[\sum_{i=1}^n\bigg(S_{j}^{h_j}[x_i]\hat{\beta}_j^{g_j}(\mathbf{x}_j)-\hat{\beta}_j^{g_j}(x_i)\bigg)\bigg]^2,
\end{align}
\noindent
and

\begin{align}\label{msehattauhat}
\widehat{MSE}_{\hat{\tau}}^{DS}=&\frac{\hat{\sigma_{\epsilon}}^2}{n^2}\sum_{i=1}^{n}\sum_{j=1}^n\bigg(S_{1}^{h_1}[x_i]S_{1}^{h_1}[x_j]^T\notag\\
&\,\,\,\,\,\,\,\,\,\,\,\,\,\,\,\,\,\,\,\,\,+S_{0}^{h_{0}}[x_i]S_{0}^{h_{0}}[x_j]^T\bigg)\notag\\
&+\bigg[\frac{1}{n}\sum_{i=1}^n\bigg(\big(S_{1}^{h_1}[x_i]\hat{\beta}_1^{g_1}(\mathbf{x}^1)-\hat{\beta}_1^{g_1}(x_i)\big)\notag\\
&\,\,\,\,\,\,\,\,\,\,\,\,\,\,\,\,\,\,\,\,\,-\big(S_{0}^{h_{0}}[x_i]\hat{\beta}_{0}^{g_{0}}(\mathbf{x}^{0})-\hat{\beta}_{0}^{g_{0}}(x_i)\big)\bigg)\bigg]^2,\\\notag
\end{align}
\noindent
where $g_1,g_{0}$ are pilot smoothing parameters selected for estimating $\beta_1$ and $\beta_{0}$ well, typically using cross-validation. The double smoothing (DS) estimation concept was utilized by \cite{hardleetal92}, although for the estimation of the entire regression function $\beta_j(\cdot)$. A difference between $\widehat{MSE}_{\bar{\hat{\beta}}_j}^{INR}$ and $\widehat{MSE}_{\bar{\hat{\beta}}_j}^{DS}$ is that the former is based on an asymptotic approximation of (\ref{mseinr}) while the double smoothing estimator targets (\ref{mseinr}) directly.

\section{Simulation study}
In this section, we study the finite sample properties of different methods for the selection of nearest neighbor type bandwidths, and in particular the resulting MSE when estimating the average causal effect $\tau$.

\subsection{Design of the study}
Data were generated according to the model
\begin{equation}\label{model.sim}
y_i=\beta_0(x_i)+\tau(x_i) z_i+\epsilon_i,\ \ \ i=1,\ldots, n,
\end{equation}
with  $x_{i} \sim \text{Uniform}(0,2\pi)$, $z_{i}|x_i \sim \text{Bernoulli}(p(x_i))$, $\epsilon_i  \sim$\\$ \text{Normal}(0,\sigma_{\epsilon}^2)$, $\tau(x_i)=\beta_1(x_i)-\beta_0(x_i)$, $\sigma_{\epsilon}^2\approx Var\big(\beta_0(x_i)+\tau(x_i) z_i\big)$, $n=100,200,500,1000$. Since $z_i$ is a Bernoulli draw dependent on $x_i$ generated from a uniform distribution, $n_1$ and $n_0$ are stochastic. Table \ref{Tab:1} and Figure \ref{fig1} display the six designs generated. Bandwidths $h_1,h_0$ considered are 40 equally spaced values within the intervals $[0.1, 1]$ for $n=100,200$ and $[0.02, 1]$ for $n=500,1000$, and, e.g., $h=0.1$ implies using 10\% of the data for the local fits. The true error variance, $\sigma_{\epsilon}^2$, is used in (\ref{msehatinr}), and (\ref{msehatinrds}) and (\ref{msehattauhat}) as well as the true propensity score, $p(x)$ in (\ref{msehatinr}). For the DS estimators in (\ref{msehatinrds}) and (\ref{msehattauhat}) the pilot bandwidths are chosen by leave-one-out cross-validation.

\begin{table*}
\footnotesize
\textit{\caption{Specification of the six designs used to generate data according to model (\ref{model.sim}).  }\label{Tab:1}}
\begin{center}
\begin{tabular}{ccc}
\hline
\newline & & \\
$Design$ &$\beta_1(x_i)$ &$\beta_0(x_i)$   \\
\hline
\hline
1 &$4\pi+5-2\pi x_i+x_i^2+5\sin(2x_i)-4\cos(x_i)$                       &$\sin(2x_i)-4\cos(x_i)+5$                     \\
2 &$4\big(x_{i}+\sin(x_{i})+\sin(2x_{i})\big)+3$                    &$2\big(x_{i}+\sin(x_{i})+\sin(2x_{i})\big)+3$                   \\
3 &$4\pi-\pi x_i+\frac{x_i^2}{2}$                     &$\pi x_i-\frac{x_i^2}{2}$                     \\
4 &$4\pi-\pi x_i+\frac{x_i^2}{2}$                    &$\pi x_i-\frac{x_i^2}{2}$                    \\
5 &$4\pi+5-2\pi x_i+x_i^2+5\sin(2x_i)-4\cos(x_i)$                  &$\sin(2x_i)-4\cos(x_i)+5$                  \\
6 &$10+x_i(2\pi-x_i)\sin(2\pi(2\pi+0.05)/(x_i+0.05))$                  &    $8+1.5\sin(2x_i-4)+6exp(-16(2x_i-2.5)^2)$              \\
\hline
\newline & & \\
$Design$  &  $\tau(x_i)$ &  $p(x_i)$  \\
\hline
\hline
1 & $4\pi-2\pi x_i+x_i^2+4\sin(2x_i)$&$[e^{-3.5+x_i}]/[1+e^{-3.5+x_i}]$  \\
2 &$ 2x_i+2\sin(x_i)+2\sin(2x_i)$ & $[e^{-3.5+x_i}]/[1+e^{-3.5+x_i}]$   \\
3 & $4\pi-2\pi x_i+x_i^2$& $[e^{-3.5+x_i}]/[1+e^{-3.5+x_i}]$  \\
4 &$4\pi-2\pi x_i+x_i^2$  &$(5\sin{2x_i}-4\cos{x_i}+4\pi-2\pi x_i+x_i^2)/11.3$   \\
5 & $4\pi-2\pi x_i+x_i^2+4\sin(2x_i)$ & $(5\sin{2x_i}-4\cos{x_i}+4\pi-2\pi x_i+x_i^2)/11.3$  \\
6 & $2+x_i(2\pi-x_i)\sin(\frac{2\pi(2\pi+0.05)}{x_i+0.05})$ & $(5\sin{2x_i}-4\cos{x_i}+4\pi-2\pi x_i+x_i^2)/11.3$  \\
&$-1.5\sin(2x_i-4)+6exp(-16(2x_i-2.5)^2)$ & \\
\hline
\end{tabular}
\end{center}
\end{table*}
{\centering{
\begin{figure*}
\begin{minipage}[b]{0.10\linewidth}
\begin{flushright}
\begin{sideways}\hspace{1.0cm}Design 6 \hspace{0.85cm}Design 5\hspace{0.85cm}Design 4\hspace{0.85cm}Design 3\hspace{0.85cm}Design 2\hspace{1cm}Design 1\end{sideways}
\end{flushright}
\end{minipage}
\begin{minipage}[b]{0.90\linewidth}
\begin{flushleft}
\centerline{\includegraphics[width=11.2cm]{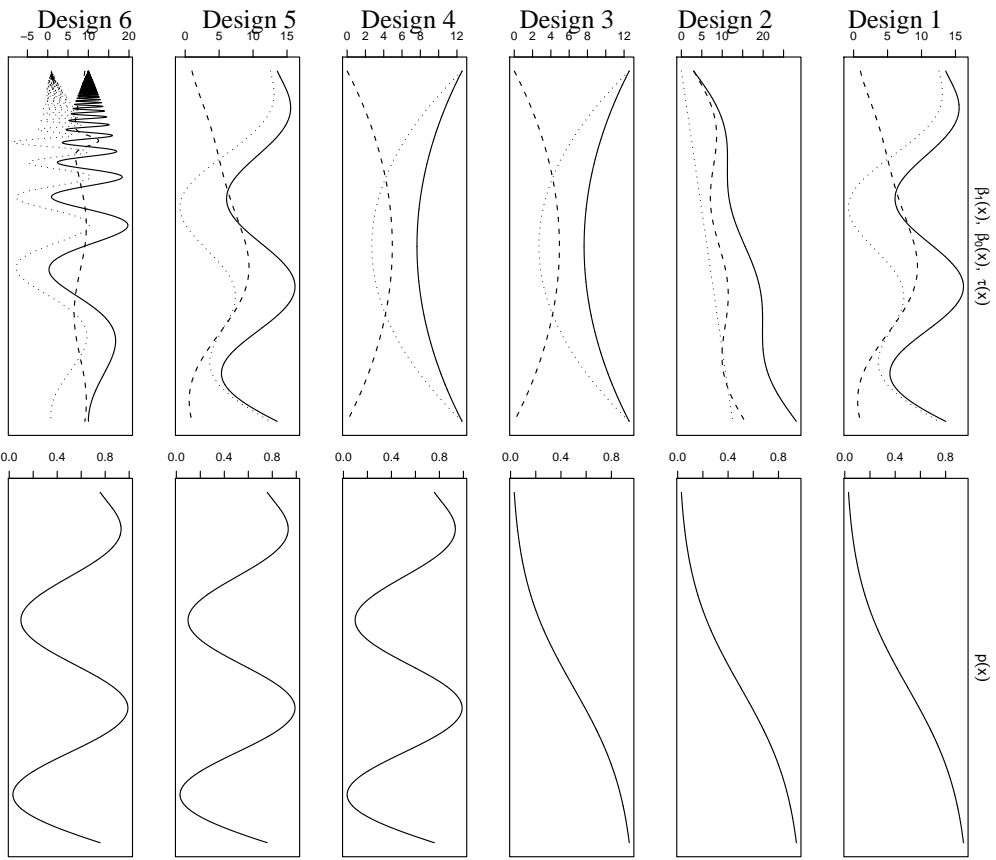}}
\end{flushleft}  
\end{minipage}
 \textit{\caption{Design 1-6 (from top to bottom) used to generate data as specified in Table \ref{Tab:1}. The first column displays $\beta_1(x_{i})$ (solid line), $\beta_0(x_{i})$ (dashed) and $\tau(x_i)$ (dotted), and the second column displays $p(x_i)$.}\label{fig1}}
\end{figure*}}}

The criteria in (\ref{msebeta}), (\ref{cv}), (\ref{mseinr}), (\ref{msetauhat}), (\ref{msehatinr}), (\ref{msehatinrds}) and (\ref{msehattauhat}) and  are computed for every bandwidth, 40 values, in the interval. For the minimizing bandwidths $\hat{\tau}^{imp}$ is computed. Due to computer time constraint, we use 200 replicates. On the other hand, we reduce noise in the simulation results by making use of the control variate method with $\hat{\tau}^{ols}$, the mean of the fitted values resulting from estimating $\tau(x)$ by ordinary least squares with correctly specified model, as control variate. If $\hat{\tau}^{ols}$ is positively correlated with $\hat{\tau}^{imp}$ then $\hat{\tau}^c=\hat{\tau}^{imp}-(\hat{\tau}^{ols}-\tau)$ has the same mean as $\hat{\tau}^{imp}$ but lower variance. For instance, for $n=1000$ such correlations varied between 0.47 and 0.95 and most of them were larger than 0.8. All computations are made in R \citep{R}. Studying bandwidth selection by simulation is computationally demanding and this study was made possible by the use of the High Performance Computing Center North (HPC2N) at Ume{\aa} University.

\subsection{Results}
\begin{figure*}
\centering
\includegraphics[scale=.56]{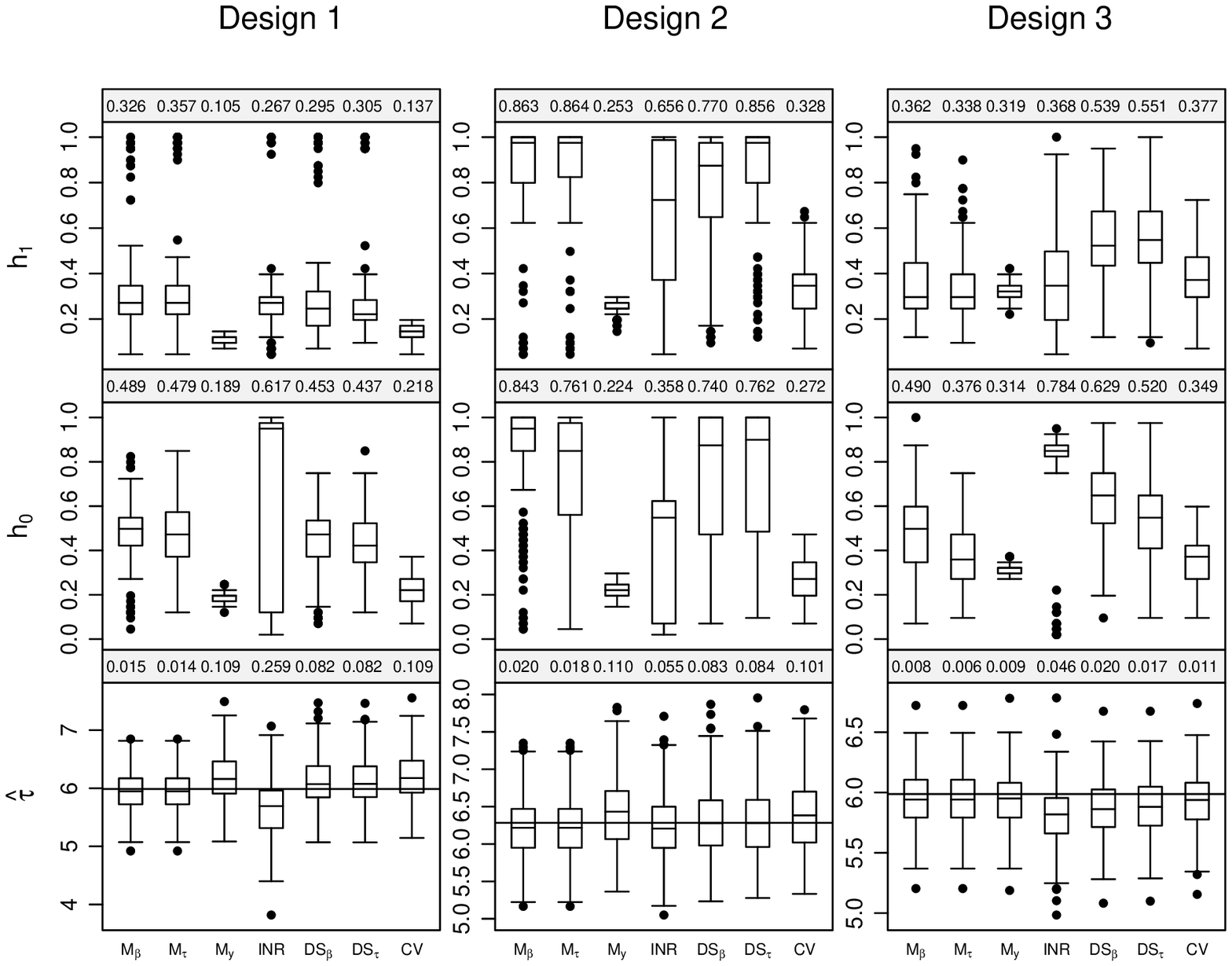}
{\caption{Design 1-3 columnwise, sample size $n=500$. Selected bandwidths and resulting $\hat\tau$ when using (\ref{mseinr}), labeled $M_{\beta}$, (\ref{msetauhat}), labeled $M_{\tau}$, (\ref{msebeta}), labeled $M_y$, (\ref{msehatinr}), labeled INR, (\ref{msehatinrds}), labeled DS$_\beta$, (\ref{msehattauhat}), labeled DS$_\tau$, and (\ref{cv}), labeled CV. Average $h$ values are given on top of the figures in the two first rows, while in the last row resulting MSEs are displayed.}\label{fig2}}
\end{figure*}

\begin{figure*}
\centering
\includegraphics[scale=.56]{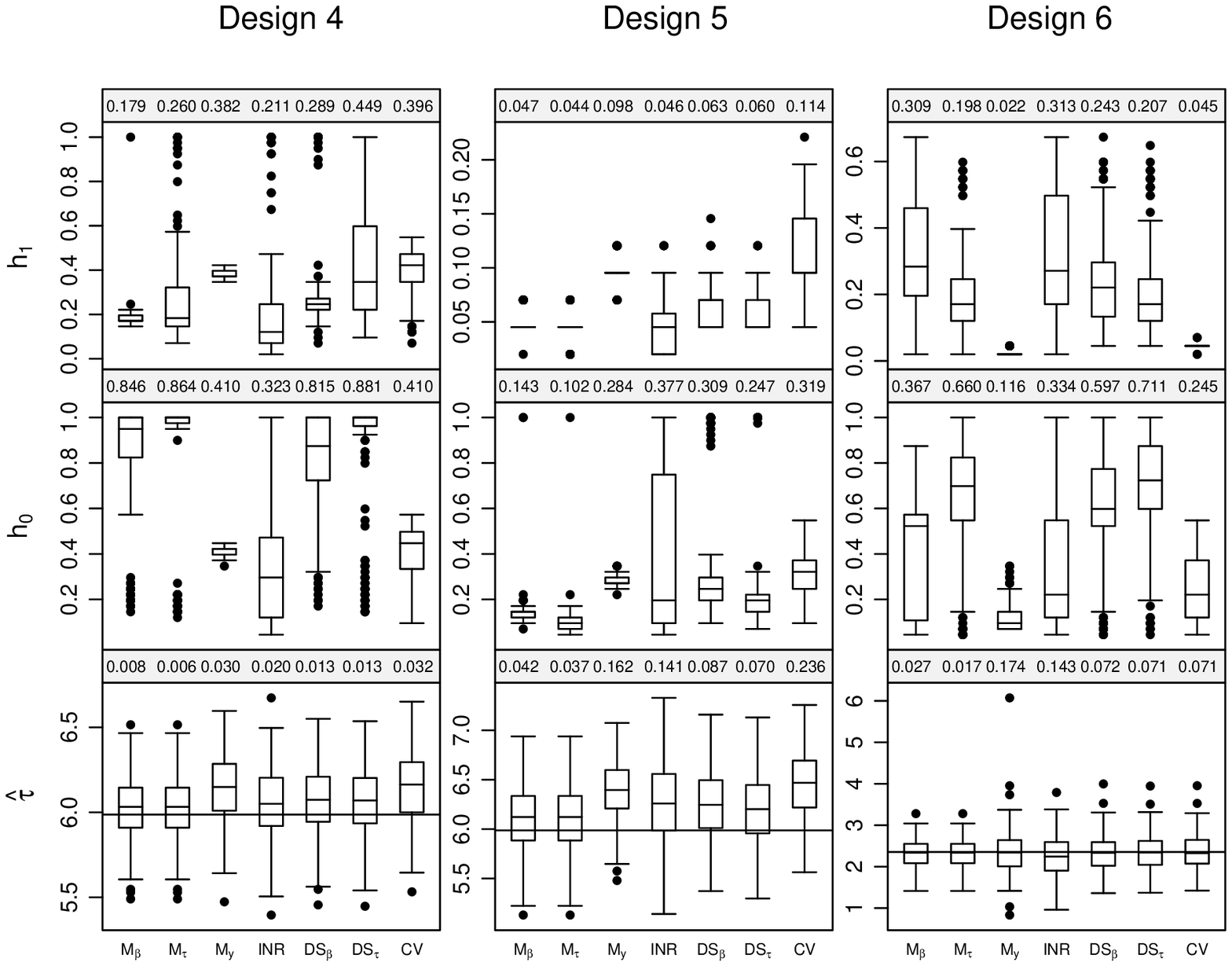}
{\caption{Design 4-6 columnwise, sample size $n=500$. Selected bandwidths and resulting $\hat\tau$ when using (\ref{mseinr}), labeled $M_{\beta}$, (\ref{msetauhat}), labeled $M_{\tau}$, (\ref{msebeta}), labeled $M_y$, (\ref{msehatinr}), labeled INR, (\ref{msehatinrds}), labeled DS$_\beta$, (\ref{msehattauhat}), labeled DS$_\tau$, and (\ref{cv}), labeled CV. Average $h$ values are given on top of the figures in the two first rows, while in the last row resulting MSEs are displayed.}\label{fig3}}
\end{figure*}

\begin{figure*}
\centering
\includegraphics[scale=.56]{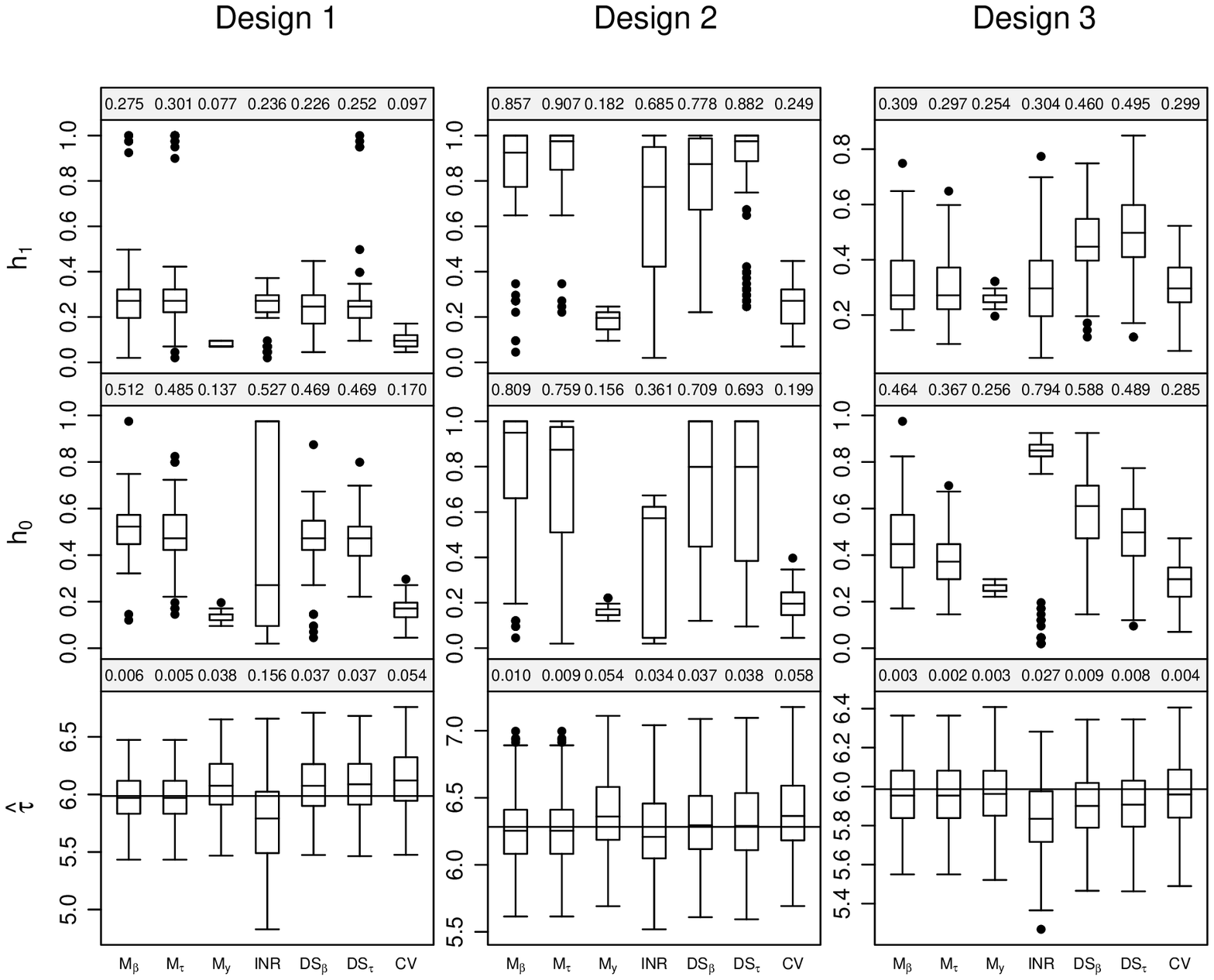}
{\caption{Design 1-3 columnwise, sample size $n=1000$. Selected bandwidths and resulting $\hat\tau$ when using (\ref{mseinr}), labeled $M_{\beta}$, (\ref{msetauhat}), labeled $M_{\tau}$, (\ref{msebeta}), labeled $M_y$, (\ref{msehatinr}), labeled INR, (\ref{msehatinrds}), labeled DS$_\beta$, (\ref{msehattauhat}), labeled DS$_\tau$, and (\ref{cv}), labeled CV. Average $h$ values are given on top of the figures in the two first rows, while in the last row resulting MSEs are displayed.}\label{fig4}}
\end{figure*}

\begin{figure*}
\centering
\includegraphics[scale=.56]{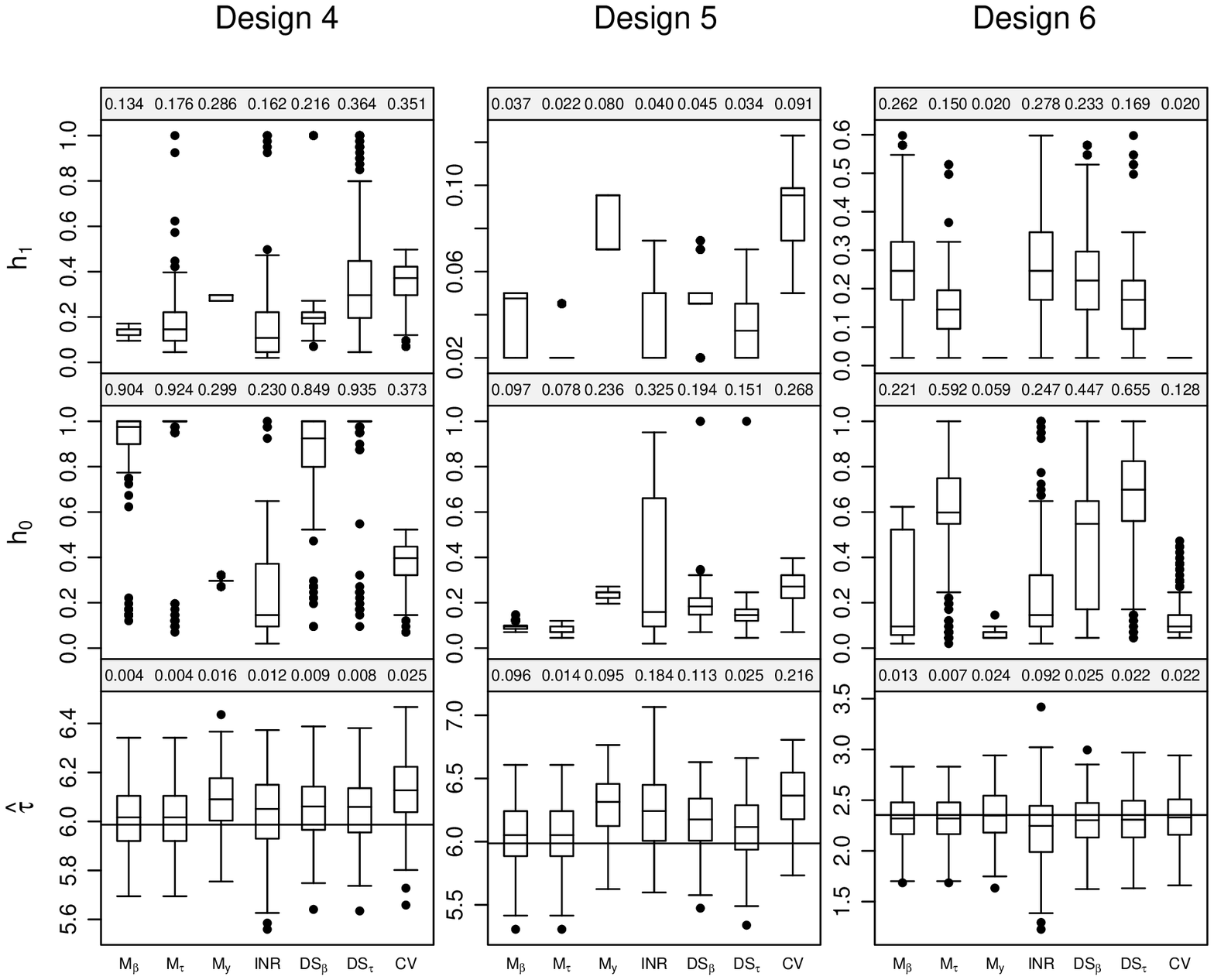}
{\caption{Design 4-6 columnwise, sample size $n=1000$. Selected bandwidths and resulting $\hat\tau$ when using (\ref{mseinr}), labeled $M_{\beta}$, (\ref{msetauhat}), labeled $M_{\tau}$, (\ref{msebeta}), labeled $M_y$, (\ref{msehatinr}), labeled INR, (\ref{msehatinrds}), labeled DS$_\beta$, (\ref{msehattauhat}), labeled DS$_\tau$, and (\ref{cv}), labeled CV. Average $h$ values are given on top of the figures in the two first rows, while in the last row resulting MSEs are displayed.}\label{fig5}}
\end{figure*}

Results for $n=500$ and 1000 are displayed in Figures \ref{fig2}-\ref{fig5} (Appendix A). More detailed results (also for $n=100,200$) are not displayed to save space but can be obtained from the authors. Note first that we can compute the smoothing parameter values minimizing (\ref{msebeta}), (\ref{mseinr}) and  (\ref{msetauhat}), labeled M$_y$, M$_{\beta}$ and M$_{\tau}$, respectively, because we know the data generating mechanisms.

We see in Figures \ref{fig2}-\ref{fig5} that the double smoothing methods introduced, (\ref{msehatinrds})  and (\ref{msehattauhat}), labeled DS$_{\beta}$ and DS$_{\tau}$ respectively, mimic quite well their target in terms of selected smoothing parameters. This is not the case for (\ref{msehatinr}), labeled INR, whose selected smoothing parameters are not in accordance with the target $M_{\beta}$. 
The results are further summarized in Tables \ref{Tab:3}-\ref{Tab:5}. Table \ref{Tab:3} summarizes MSE results given in Figures \ref{fig2}-\ref{fig5} (for $n=500,1000$) for the theoretical criteria M$_{\beta}$, M$_{\tau}$ and M$_y$, by indicating which criterion yielded lowest MSE for the estimation of $\tau$. We see that M$_\tau$ always results in smallest MSE, which is, in most cases, significantly smaller than the second smallest MSE (achieved by M$_\beta$ except for Design 3, $n=200$, and Design 5, $n=1000$). Both M$_{\tau}$ and M$_{\beta}$ result in significantly smaller MSE than M$_y$ in all cases but three (Design 3, $n=200, 1000$, Design 5, $n=1000$).
Table \ref{Tab:5} gives information on MSE (similar to Table \ref{Tab:3}), where comparisons are made between the data-driven criteria DS$_{\beta}$, DS$_{\tau}$, INR and CV.   
We see that double smoothing does not always yields lowest MSE, although CV is most often outperformed by the methods targeting the estimation of functional averages (DS and INR $-$for design 2 where INR performed best, CV was also outperformed by DS). 

Finally, note that the propensity scores used in the designs of this study are rather extreme in the sense that they may yield probabilities near zero and one. We have also run these experiments by damping these propensity scores to let them vary only between 0.2 and 0.8. The results where similar qualitatively with double smoothing performing better.

\begin{table}[h!]
\footnotesize
\begin{center}
{\caption{MSE comparison: The table displays the method yielding lowest MSE among M$_{\beta}$, M$_{\tau}$ and M$_{y}$. Stars indicate that the method has significantly lower MSE than the next best method, with ``*'' for a 5\% level test and ``**'' for a 1\% level test.}\label{Tab:3}}
\begin{tabular}{cllll}
\hline
\newline
 & \multicolumn{4}{c}{}  \\
$Design$      & \multicolumn{4}{c}{Minimum MSE obtained by}  \\
\hline
\hline
 & \multicolumn{4}{c}{$n$} \\
\hline
   &$100$ &$200$  &$500$ &$1000$\\
\hline
1    &M$_{\tau}$ &M$_{\tau}^{*}$ &M$_{\tau}$  &M$_{\tau}^{**}$ \\
2    &M$_{\tau}^{**}$ &M$_{\tau}$ &M$_{\tau}^{**}$  &M$_{\tau}$ \\
3    &M$_{\tau}^{**}$ &M$_{\tau}$ &M$_{\tau}^{**}$  &M$_{\tau}^{**}$ \\
4    &M$_{\tau}$ &M$_{\tau}^{**}$ &M$_{\tau}^{**}$  &M$_{\tau}^{**}$ \\
5    &M$_{\tau}^{*}$ &M$_{\tau}$ &M$_{\tau}^{**}$  &M$_{\tau}^{**}$ \\
6    &M$_{\tau}$ &M$_{\tau}$ &M$_{\tau}^{**}$  &M$_{\tau}^{**}$ \\
 \hline
\end{tabular}
\end{center}
\end{table}

\begin{table}[h!]
\footnotesize
\begin{center}
{\caption{MSE comparison: The table displays the method yielding lowest MSE among DS$_{\beta}$, DS$_{\tau}$, INR and CV. Stars indicate that the method has significantly lower MSE than the next best method, with ``*'' for a 5\% level test and ``**'' for a 1\% level test.}\label{Tab:5}}
\begin{tabular}{cllll}
\hline
\newline
 & \multicolumn{4}{c}{}  \\
$Design$      & \multicolumn{4}{c}{Minimum MSE obtained by}  \\
\hline
\hline
 & \multicolumn{4}{c}{$n$} \\
\hline
   &$100$ &$200$  &$500$ &$1000$\\
\hline
1    &DS$_{\beta}$ &DS$_{\beta}^{**}$ &DS$_{\tau}$  &DS$_{\tau}$ \\
2    &INR$^{**}$ &INR$^{*}$ &INR & INR\\   
3    &CV$^{*}$ &CV$^{**}$ &CV$^{**}$& CV$^{**}$\\   
4    &DS$_{\tau}$ &~DS$_{\tau}^{**}$ &DS$_{\tau}$ & DS$_{\tau}^{*}$\\   
5    &DS$_{\tau}^{**}$ &DS$_{\tau}^{**}$ & DS$_{\tau}^{**}$& DS$_{\tau}^{**}$\\   
6    & CV&DS$_{\beta}$ &DS$_{\tau}$ & DS$_{\tau}$\\   
 \hline
\end{tabular}
\end{center}
\end{table}

\section{Conclusion}
In this paper we have proposed double smoothing methods for selecting smoothing parameters that target the estimation of functional averages where the latter are average causal effects of interest. In our numerical experiments cross-validation is often outperformed by double smoothing as we expected since the latter criterion is optimized for the estimation of functions underlying the average causal effect, and not the average itself. The methods proposed and studied here have large applicability, and are, for instance, straightforward to adapt to non-parametric estimators based on instruments as those introduced in \cite{FR:07}.

\section*{Acknowledgments}
We are grateful to Yanyuan Ma and Sara Sj\"ostedt-de Luna for comments that have helped us to improve the paper. We acknowledge the financial support of the Swedish Research Council through the Swedish Initiative for Research on Microdata in the Social and Medical Sciences (SIMSAM), the Ageing
and Living Condition Program and grant 70246501.

\section*{Appendix}
\subsection{Figures with results}
\null
\vskip 2cm

\clearpage

\clearpage

\subsection{Asymptotics}
In order to derive the results of Section 3.2 we focus on local linear regression with constant bandwidth. 
We use further the following assumptions.\\
(A1) The kernel $K$ is a compactly supported, bounded kernel such that $\int u^2 K(u) du\neq 0$. In addition, all odd-order moments of $K$ vanish, that is $\int u^lK(u)du=0$ for all nonnegative odd integers $l$.\\
(A2) The covariate $x$ has density $f$. The point $\tilde x$ is in the interior of supp$(f)=\{x\in \mathbb{R}:f(x)>0\}$. At $\tilde x$, $f$ is continuously differentiable and all second-order derivatives of $\beta_j$, $j=0,1$, are continuous. \\
(A3) For $j=0,1$, $h_j\rightarrow 0$ and $nh_j \rightarrow \infty$ as $n\rightarrow \infty$ .\\
\noindent
\\
We have
 \begin{align*}
MSE_{\bar{\hat{\beta}}_j}=&\frac{1}{n^2}\sum_{i=1}^nVar(\hat{\beta}_{j}^{h_j}(x_i)|\mathbf{x})\notag\\
&+\frac{1}{n^2}\sum_{\begin{subarray}{l} i=1\\i\neq l\end{subarray}}^n\sum_{l=1}^nCov(\hat{\beta}_{j}^{h_j}(x_i),\hat{\beta}_{j}^{h_j}(x_l)|\mathbf{x})\\
&+\bigg[\frac{1}{n}\sum_{i=1}^nE(\hat{\beta}_{j}^{h_j}(x_i)-\beta_j(x_i)|\mathbf{x})\bigg]^2.\\
\end{align*}
Under (A1)-(A2) for $\tilde x=x_i$, \citet[Thm 2.1]{ruwa94}  states that
\begin{equation}\label{varbeta}
Var(\hat{\beta}_{j}^{h_j}(x_i)|\mathbf{x})=\frac{\sigma_{\epsilon}^2}{n_jh_j}f_j(x_i)^{-1}\int K(u)^2du\{1+o_p(1)\}
\end{equation}
and
\begin{align}\label{biasbeta}
E(\hat{\beta}_{j}^{h_j}(x_i)&-\beta_j(x_i)|\mathbf{x})\notag\\
&=\frac{h_j^2}{2}\beta_j^{(2)}(x_i)\int u^2K(u)du\{1+o_p(1)\},
\end{align}
\noindent
where $f_j(x_i)=f(x_i|z_i=j)=\frac{f(x_i)\Pr(z_i=j|x_i)}{\Pr(z_i=j)}$. It follows from (\ref{varbeta}) and the fact that $n_j=(-1)^{j+1}\sum_{i=1}^n z_i+n(1-j)$ that
\begin{align}\label{meanvarbeta}
&\frac{1}{n^2}\sum_{i=1}^nVar(\hat{\beta}_{j}^{h_j}(x_i)|\mathbf{x})\notag\\
&\,\,=\frac{1}{n^2}\sum_{i=1}^n\frac{\sigma_{\epsilon}^2}{n_jh_j}f_j(x_i)^{-1}\int K(u)^2du\{1+o_p(1)\}\notag\\
&\,\,=\frac{\sigma_{\epsilon}^2}{n^2h_j}\int K(u)^2du\int\frac{1}{\Pr(z_i=j|x)}dx+o_p(n^{-2}h_j^{-1}).\\
\notag
\end{align}
\noindent
Using (\ref{biasbeta}) we have
\begin{align}\label{biasmeanbeta}
\frac{1}{n}\sum_{i=1}^n&E(\hat{\beta}_{j}^{h_j}(x_i)-\beta_j(x_i)|\mathbf{x})\notag\\
=&\frac{1}{n}\sum_{i=1}^n\frac{h_j^2}{2}\beta_j^{(2)}(x_i)\int u^2K(u)du\{1+o_p(1)\}\notag\\
=&\frac{h_j^2}{2}\int \beta_j^{(2)}(x)f(x)dx\int u^2K(u)du+o_p(h_j^2).\\\notag
\end{align}
\noindent
Now,
\begin{align}\label{cov}
&\frac{1}{n^2}\sum_{\begin{subarray}{l} i=1\\i\neq l\end{subarray}}^n\sum_{l=1}^nCov(\hat{\beta}_j^{h_j}(x_i),\hat{\beta}_j^{h_j}(x_l)|\mathbf{x})\notag\\
&\,\,=\frac{1}{n^2}\sum_{\begin{subarray}{l} i=1\\i\neq l\end{subarray}}^n\sum_{l=1}^nE\big(S_{j}^{h_j}[x_i]\mathbf{\epsilon}_j\mathbf{\epsilon}_j^{T}S_{j}^{h_j}[x_l]^{T}|\mathbf{x}\big)\notag\\
&\,\,=\frac{1}{n^2}\sum_{\begin{subarray}{l} i=1\\i\neq l\end{subarray}}^n\sum_{l=1}^n\sigma_{\epsilon}^2\mathbf{e}_1^T (n_j^{-1}\mathbf{X}_{i}^{jT}\mathbf{W}_i^{h_j}\mathbf{X}_{i}^j)^{-1}n_j^{-2}\notag\\
&\,\,\,\,\,\,\,\,\,\,\,\,\,\,\,\,\,\,\,\,\,\,\,\,\,\,\,\,\,\,\,\,\times\mathbf{X}_{i}^{jT}\mathbf{W}_i^{h_j}\mathbf{W}_l^{h_j}\mathbf{X}_{l}^{j}(n_j^{-1}\mathbf{X}_{l}^{jT}\mathbf{W}_l^{h_j}\mathbf{X}_{l}^{j})^{-1}\mathbf{e}_1.\notag\\
\end{align}
\noindent
According to  \citet[eq. (2.11)]{ruwa94} 
\begin{align}
(n_j^{-1}&\mathbf{X}_{i}^{jT}\mathbf{W}_i^{h_j}\mathbf{X}_{i}^{j})^{-1}\notag\\
&=
\left( \begin{array}{cc}
f_j(x_i)^{-1}+o_p(1) & -\frac{f_j^{(1)}(x_i)}{f_j(x_i)^2}+o_p(1) \\
-\frac{f_j^{(1)}(x_i)}{f_j(x_i)^2}+o_p(1) &\{\int u^2K(u)duf_j(x_i)h_j^2\}^{-1}+o_p(h_j^{-2})
\end{array} \right ).\notag
\end{align}
\noindent
Noting that
\begin{align}
\{n_j^{-2}&\mathbf{X}_{i}^{jT}\mathbf{W}_i^{h_j}\mathbf{W}_l^{h_j}\mathbf{X}_{l}^{j}\}_{11}\notag\\
&=\frac{1}{n_j^2h_j^2}\sum_{k=1}^{n_j}K\bigg(\frac{x_k-x_i}{h_j}\bigg)K\bigg(\frac{x_k-x_l}{h_j}\bigg),\notag
\end{align}

\begin{align}
\{n_j^{-2}&\mathbf{X}_{i}^{jT}\mathbf{W}_i^{h_j}\mathbf{W}_l^{h_j}\mathbf{X}_{l}^{j}\}_{12}\notag\\
&=\frac{1}{n_j^2h_j^2}\sum_{k=1}^{n_j}K\bigg(\frac{x_k-x_i}{h_j}\bigg)K\bigg(\frac{x_k-x_l}{h_j}\bigg)(x_k-x_l),  \notag
\end{align}

\begin{align}
\{n_j^{-2}&\mathbf{X}_{i}^{jT}\mathbf{W}_i^{h_j}\mathbf{W}_l^{h_j}\mathbf{X}_{l}^{j}\}_{21}\notag\\
&=\frac{1}{n_j^2h_j^2}\sum_{k=1}^{n_j}K\bigg(\frac{x_k-x_i}{h_j}\bigg)K\bigg(\frac{x_k-x_l}{h_j}\bigg)(x_k-x_i),  \notag
\end{align}
and
\begin{align}
\{n_j^{-2}&\mathbf{X}_{i}^{jT}\mathbf{W}_i^{h_j}\mathbf{W}_l^{h_j}\mathbf{X}_{l}^{j}\}_{22}\notag\\
&=\frac{1}{n_j^2h_j^2}\sum_{k=1}^{n_j}K\bigg(\frac{x_k-x_i}{h_j}\bigg)K\bigg(\frac{x_k-x_l}{h_j}\bigg)(x_k-x_i)(x_k-x_l). \notag
\end{align}
It follows that the covariance in (\ref{cov}) can be written as 
\begin{align}
&\frac{1}{n^2}\sum_{\begin{subarray}{l} i=1\\i\neq l\end{subarray}}^n\sum_{l=1}^nCov(\hat{\beta}_j^{h_j}(x_i),\hat{\beta}_j^{h_j}(x_l)|\mathbf{x})\notag\\
=&\frac{\sigma_{\epsilon}^2}{n^2}\sum_{\begin{subarray}{l} i=1\\i\neq l\end{subarray}}^n\sum_{l=1}^n\frac{1}{f_j(x_i)f_j(x_l)}\frac{1}{n_j^2h_j^2}\sum_{k=1}^{n_j}K\bigg(\frac{x_k-x_i}{h_j}\bigg)\notag\\
& \,\,\,\,\,\,\,\,\,\,\,\,\,\,\,\,\,\,\,\,\,\,\,\,\,\,\,\,\,\,\,\times K\bigg(\frac{x_k-x_l}{h_j}\bigg)\{1+o_p(1)\}\notag\\
&-\frac{\sigma_{\epsilon}^2}{n^2}\sum_{\begin{subarray}{l} i=1\\i\neq l\end{subarray}}^n\sum_{l=1}^n\frac{f_j^{(1)}(x_i)}{f_j(x_i)^2f_j(x_l)} \frac{1}{n_j^2h_j^2}\sum_{k=1}^{n_j}K\bigg(\frac{x_k-x_i}{h_j}\bigg)\notag\\
& \,\,\,\,\,\,\,\,\,\,\,\,\,\,\,\,\,\,\,\,\,\,\,\,\,\,\,\,\,\,\,\times K\bigg(\frac{x_k-x_l}{h_j}\bigg)(x_k-x_i)\{1+o_p(1)\}\notag\\
&-\frac{\sigma_{\epsilon}^2}{n^2}\sum_{\begin{subarray}{l} i=1\\i\neq l\end{subarray}}^n\sum_{l=1}^n\frac{f_j^{(1)}(x_l)}{f_j(x_l)^2f_j(x_i)}\frac{1}{n_j^2h_j^2}\sum_{k=1}^{n_j}K\bigg(\frac{x_k-x_i}{h_j}\bigg)\notag\\
& \,\,\,\,\,\,\,\,\,\,\,\,\,\,\,\,\,\,\,\,\,\,\,\,\,\,\,\,\,\,\,\times K\bigg(\frac{x_k-x_l}{h_j}\bigg)(x_k-x_l)\{1+o_p(1)\}\notag\\
&+\frac{\sigma_{\epsilon}^2}{n^2}\sum_{\begin{subarray}{l} i=1\\i\neq l\end{subarray}}^n\sum_{l=1}^n\frac{f_j^{(1)}(x_i)f_j^{(1)}(x_l)}{f_j(x_i)^2f_j(x_l)^2}\frac{1}{n_j^2h_j^2}\sum_{k=1}^{n_j}K\bigg(\frac{x_k-x_i}{h_j}\bigg)\notag\\
& \,\,\,\,\,\,\,\,\,\,\,\,\,\,\,\,\,\,\,\,\,\,\,\,\,\,\,\,\,\,\,\times K\bigg(\frac{x_k-x_l}{h_j}\bigg)(x_k-x_i)(x_k-x_l)\notag\\
& \,\,\,\,\,\,\,\,\,\,\,\,\,\,\,\,\,\,\,\,\,\,\,\,\,\,\,\,\,\,\,\times\{1+o_p(1)\}.\notag\\\notag
\end{align}
\noindent
Now,
\begin{align}\label{i}
&\frac{\sigma_{\epsilon}^2}{n^2}\sum_{\begin{subarray}{l} i=1\\i\neq l\end{subarray}}^n\sum_{l=1}^n\frac{1}{f_j(x_i)f_j(x_l)}\frac{1}{n_j^2h_j^2}\sum_{k=1}^{n_j}K\bigg(\frac{x_k-x_i}{h_j}\bigg)\notag\\
& \,\,\,\,\,\,\,\,\,\,\,\,\,\,\,\,\,\,\,\,\,\,\,\,\,\,\,\,\,\,\, \times K\bigg(\frac{x_k-x_l}{h_j}\bigg)\{1+o_p(1)\}\notag\\
&=\frac{\sigma_{\epsilon}^2}{n_j^2}\frac{(n-1)}{n}\sum_{k=1}^{n_j}\bigg[\int \frac{1}{h_j}K\bigg(\frac{x_k-x_i}{h_j}\bigg)\frac{f(x_i)}{f_j(x_i)}\bigg]\notag\\
& \,\,\,\,\,\,\,\,\,\,\,\,\,\,\,\,\,\,\,\,\,\,\,\,\,\,\,\,\,\,\, \times\bigg[\int \frac{1}{h_j}K\bigg(\frac{x_k-x_l}{h_j}\bigg)\frac{f(x_l)}{f_j(x_l)}dx_l\bigg]\notag\\
& \,\,\,\,\,\,\,\,\,\,\,\,\,\,\,\,\,\,\,\,\,\,\,\,\,\,\,\,\,\,\, \times\{1+o_p(1)\}\notag\\
&=\frac{\sigma_{\epsilon}^2}{n_j^2}\frac{(n-1)}{n}\sum_{k=1}^{n_j}\bigg[\int K(-u)\frac{f(x_k+h_ju)}{f_j(x_k+h_ju)}du\bigg]\notag\\
& \,\,\,\,\,\,\,\,\,\,\,\,\,\,\,\,\,\,\,\,\,\,\,\,\,\,\,\,\,\,\, \times\bigg[\int K(-u)\frac{f(x_k+h_ju)}{f_j(x_k+h_ju)}du\bigg]\{1+o_p(1)\}\notag\\
&=\frac{\sigma_{\epsilon}^2}{n}\bigg[\int \frac{f(x)}{\Pr(z=j|x)}dx\bigg]+o_p(n^{-1})\\\notag
\end{align}
and

\begin{align}\label{ii}
&-\frac{\sigma_{\epsilon}^2}{n^2}\sum_{\begin{subarray}{l} i=1\\i\neq l\end{subarray}}^n\sum_{l=1}^n\frac{f_j^{(1)}(x_i)}{f_j(x_i)^2f_j(x_l)} \frac{1}{n_j^2h_j^2}\sum_{k=1}^{n_j}K\bigg(\frac{x_k-x_i}{h_j}\bigg)\notag\\
& \,\,\,\,\,\,\,\,\,\,\,\,\,\,\,\,\,\,\,\,\,\,\,\,\,\,\,\,\,\,\, \times K\bigg(\frac{x_k-x_l}{h_j}\bigg)(x_k-x_i)\{1+o_p(1)\}\notag\\
=&-\frac{\sigma_{\epsilon}^2(n-1)}{\Pr(z=j)n^2}\frac{1}{(n-1)}\sum_{\begin{subarray}{l} i=1\\i\neq l\end{subarray}}^n\frac{f_j^{(1)}(x_i)}{f_j(x_i)^2}\frac{1}{n_j}\sum_{k=1}^{n_j}\frac{1}{h_j}K\bigg(\frac{x_k-x_i}{h_j}\bigg)\notag\\
& \,\,\,\,\,\,\,\,\,\,\,\,\,\,\,\,\,\,\,\,\,\,\,\,\,\,\,\,\,\,\,\times (x_k-x_i)\bigg[\int \frac{1}{h_j}K\bigg(\frac{x_k-x_l}{h_j}\bigg)\frac{f(x_l)}{f_j(x_l)}\bigg] \{1+o_p(1)\}\notag\\
=&-\frac{\sigma_{\epsilon}^2(n-1)}{\Pr(z=j)n^2}\frac{1}{(n-1)}\sum_{\begin{subarray}{l} i=1\\i\neq l\end{subarray}}^n\frac{f_j^{(1)}(x_i)}{f_j(x_i)^2}\frac{1}{n_j}\sum_{k=1}^{n_j}\frac{1}{h_j}K\bigg(\frac{x_k-x_i}{h_j}\bigg)\notag\\
& \,\,\,\,\,\,\,\,\,\,\,\,\,\,\,\,\,\,\,\,\,\,\,\,\,\,\,\,\,\,\,\times(x_k-x_i)\bigg[\int K(-u)\frac{f(x_k+h_ju)}{f_j(x_k+h_ju)}du\bigg]\{1+o_p(1)\}\notag\\
=&-\frac{\sigma_{\epsilon}^2(n-1)}{\Pr(z=j)n^2}\frac{1}{(n-1)}\sum_{\begin{subarray}{l} i=1\\i\neq l\end{subarray}}^n\frac{f_j^{(1)}(x_i)}{f_j(x_i)^2}\bigg[\int \frac{1}{h_j}K\bigg(\frac{x_k-x_i}{h_j}\bigg)\notag \\
& \,\,\,\,\,\,\,\,\,\,\,\,\,\,\,\,\,\,\,\,\,\,\,\,\,\,\,\,\,\,\, \times (x_k-x_i)f(x_k)dx_k\bigg]\{1+o_p(1)+O(h_j)\}\notag\\
=&-\frac{\sigma_{\epsilon}^2(n-1)}{\Pr(z=j)n^2}\frac{1}{(n-1)}\sum_{\begin{subarray}{l} i=1\\i\neq l\end{subarray}}^n\frac{f_j^{(1)}(x_i)}{f_j(x_i)^2}\bigg[\int u^2K(u)h_j^2\frac{f(x_i+h_ju)}{h_ju}du\bigg]\notag\\
& \,\,\,\,\,\,\,\,\,\,\,\,\,\,\,\,\,\,\,\,\,\,\,\,\,\,\,\,\,\,\, \times\{1+o_p(1)+O(h_j)\}\notag\\
=&-\frac{\sigma_{\epsilon}^2h_j^2}{\Pr(z=j)n}\int u^2K(u)du\bigg[\int \frac{f_j^{(1)}(x_i)f^{(1)}(x_i)}{f_j(x_i)^2}f(x_i)dx_i\bigg]\notag\\
&+o_p(n^{-1}h_j^2)\notag\\
=&-\frac{\sigma_{\epsilon}^2h_j^2}{n}\int u^2K(u)du\bigg[\int \frac{f^{(1)}(x)^2}{f(x)\Pr(z=j|x)}\notag\\
& \,\,\,\,\,\,\,\,\,\,\,\,\,\,\,\,\,\,\,\,\,\,\,\,\,\,\,\,\,\,\,+\frac{f^{(1)}(x)P^{(1)}(z=j|x)}{\Pr(z=j|x)^2}dx\bigg]+o_p(n^{-1}h_j^2).\\\notag
\end{align}
\noindent
Analogously,
\begin{align}\label{iii}
-\frac{\sigma_{\epsilon}^2}{n^2}\sum_{\begin{subarray}{l} i=1\\i\neq l\end{subarray}}^n\sum_{l=1}^n&\frac{f_j^{(1)}(x_l)}{f_j(x_l)^2f_j(x_i)}\frac{1}{n_j^2h_j^2}\sum_{k=1}^{n_j}K\bigg(\frac{x_k-x_i}{h_j}\bigg)K\bigg(\frac{x_k-x_l}{h_j}\bigg)\notag\\
& \,\,\,\,\,\,\,\,\,\,\,\,\,\,\,\,\,\,\,\,\,\,\,\,\,\,\,\,\,\,\, \times(x_k-x_l)\{1+o_p(1)\}\notag\\
=&-\frac{\sigma_{\epsilon}^2h_j^2}{n}\int u^2K(u)du\bigg[\int \frac{f^{(1)}(x)^2}{f(x)\Pr(z=j|x)}\notag\\
&  \,\,\,\,\,\,\,\,\,\,\,\,\,\,\,\,\,\,\,\,\,\,\,\,\,\,\,\,\,\,\,+\frac{f^{(1)}(x)P^{(1)}(z=j|x)}{\Pr(z=j|x)^2}dx\bigg]\notag\\
&+o_p(n^{-1}h_j^2).\\\notag
\end{align}
\noindent
Finally, 
\begin{align}\label{iv}
&\frac{\sigma_{\epsilon}^2}{n^2}\sum_{\begin{subarray}{l} i=1\\i\neq l\end{subarray}}^n\sum_{l=1}^n\frac{f_j^{(1)}(x_i)f_j^{(1)}(x_l)}{f_j(x_i)^2f_j(x_l)^2}\frac{1}{n_j^2h_j^2}\sum_{k=1}^{n_j}K\bigg(\frac{x_k-x_i}{h_j}\bigg)\notag\\
& \times K\bigg(\frac{x_k-x_l}{h_j}\bigg)(x_k-x_i)(x_k-x_l)\{1+o_p(1)\}.\notag\\
=&\frac{\sigma_{\epsilon}^2}{\Pr(z=j)n^3}\sum_{\begin{subarray}{l} i=1\\i\neq l\end{subarray}}^n\sum_{l=1}^n\frac{f_j^{(1)}(x_i)f_j^{(1)}(x_l)}{f_j(x_i)^2f_j(x_l)^2}\bigg[\int \frac{1}{h_j^2}K\bigg(\frac{x_k-x_i}{h_j}\bigg)\notag \\
& \,\,\,\,\,\,\,\,\,\,\,\,\,\,\,\,\,\,\,\,\,\,\,\,\,\,\,\,\,\, \times K\bigg(\frac{x_k-x_l}{h_j}\bigg)(x_k-x_i)(x_k-x_l)f_j(x_k)dx_k\bigg]\notag\\
&\times \{1+o_p(1)\}\notag\\
=&\frac{\sigma_{\epsilon}^2}{\Pr(z=j)n^3}\sum_{\begin{subarray}{l} i=1\\i\neq l\end{subarray}}^n\sum_{l=1}^n\frac{f_j^{(1)}(x_i)f_j^{(1)}(x_l)}{f_j(x_i)^2f_j(x_l)^2}\bigg[\int K(u)\notag\\
&\,\,\,\,\,\,\,\,\,\,\,\,\,\,\,\,\,\,\,\,\,\,\,\,\,\,\,\,\,\,\times K\bigg(\frac{x_i+h_ju-x_l}{h_j}\bigg)u(x_i+h_ju-x_l)\notag\\
&\,\,\,\,\,\,\,\,\,\,\,\,\,\,\,\,\,\,\,\,\,\,\,\,\,\,\,\,\,\, \times f_j(x_i+h_ju)du\bigg]\{1+o_p(1)\}\notag\\
=&0.\\\notag
\end{align}
\noindent
It follows from (\ref{i})-(\ref{iv}) that
\begin{align}\label{meancov}
&\frac{1}{n^2}\sum_{\begin{subarray}{l} i=1\\i\neq l\end{subarray}}^n\sum_{l=1}^nCov(\hat{\beta}_j^{h_j}(x_i),\hat{\beta}_j^{h_j}(x_l)|\mathbf{x})\notag\\
&\,\, =\frac{\sigma_{\epsilon}^2}{n}\bigg[\int \frac{f(x_k)}{\Pr(z_k=j|x_k)}dx_k\bigg]+o_p(n^{-1})\notag\\
&\,\,\,-\frac{2\sigma_{\epsilon}^2h_j^2}{n}\int u^2K(u)du \int\bigg( \frac{f^{(1)}(x)^2}{f(x)\Pr(z=j|x)}\notag \\
&\,\,\,\,\,\,\,\,\,+\frac{f^{(1)}(x)P^{(1)}(z=j|x)}{\Pr(z=j|x)^2}\bigg)dx+o_p(n^{-1}h_j^2).\\\notag
\end{align}

\noindent
It follows from (\ref{meanvarbeta}) and (\ref{meancov}) that
\begin{align}\label{varmeanbeta}
&Var\bigg(\frac{1}{n}\sum_{i=1}^n\hat{\beta}_j^{h_j}(x_i)\Big|\mathbf{x}\bigg)\notag\\
&\,\, =\frac{1}{n^2}\sum_{i=1}^nVar(\hat{\beta}_j^{h_j}(x_i)|\mathbf{x})+\frac{1}{n^2}\sum_{\begin{subarray}{l} i=1\\i\neq l\end{subarray}}^n\sum_{l=1}^nCov(\hat{\beta}_j^{h_j}(x_i),\hat{\beta}_j^{h_j}(x_l)|\mathbf{x})\notag\\
&\,\, =\frac{\sigma_{\epsilon}^2}{n^2h_j}\int K(u)^2du\int\frac{1}{\Pr(z=j|x)}dx\notag\\
&\,\, \,\,\,\,\,+\frac{\sigma_{\epsilon}^2}{n}\bigg[\int \frac{f(x)}{\Pr(z=j|x)}dx\bigg]\notag\\
&\,\, \,\,\,\,\,-\frac{2\sigma_{\epsilon}^2h_j^2}{n}\int u^2K(u)du \int\bigg( \frac{f^{(1)}(x)^2}{f(x)\Pr(z=j|x)}\notag \\
&\,\, \,\,\,\,\,\,\, \,\,\,\,\,\,\, \,\,\,\,\,+\frac{f^{(1)}(x)P^{(1)}(z=j|x)}{\Pr(z=j|x)^2}\bigg)dx+o_p(n^{-2}h_j^{-1}+n^{-1}+n^{-1}h_j^2).\\\notag
\end{align}
\noindent
Hence, from (\ref{varmeanbeta}) and (\ref{biasmeanbeta}),
\begin{align}
&MSE\bigg(\frac{1}{n}\sum_{i=1}^n\hat{\beta}_j^{h_j}(x_i)\Big|\mathbf{x}\bigg)\notag\\
&\,\, =\frac{\sigma_{\epsilon}^2}{n^2h_j}\int K(u)^2du\int\frac{1}{\Pr(z=j|x)}dx+\frac{\sigma_{\epsilon}^2}{n}\bigg[\int \frac{f(x)}{\Pr(z=j|x)}dx\bigg]\notag\\
&-\frac{2\sigma_{\epsilon}^2h_j^2}{n}\int u^2K(u)du\notag \\
&\times\int\bigg( \frac{f^{(1)}(x)^2}{f(x)\Pr(z=j|x)}+\frac{f^{(1)}(x)P^{(1)}(z=j|x)}{\Pr(z=j|x)^2}\bigg)dx\notag \\
&+\frac{h_j^4}{4}\bigg[\int \beta_j^{(2)}(x)f(x)dx\bigg]^2\bigg[\int u^2K(u)du\bigg]^2\notag \\
&+o_p(n^{-2}h_j^{-1}+n^{-1}+n^{-1}h_j^2+h_j^4).\notag \\
\notag
\end{align}
\noindent
Finally, (\ref{asymnorminra})-(\ref{asymnorminrd}) follows from (\ref{condbias.eq}) and (\ref{condvar.eq}). By the weak law of large numbers we can write
$$\frac{1}{n}\sum_{i=1}^n\beta_j(x_i)-E(\beta_j(x_i))=o_p(1). $$
Combined with (\ref{condbias.eq}) we thus have
$$E\left(\frac{1}{n}\sum_{i=1}^n\hat{\beta}_j^{h_j}(x_i)\Big|\mathbf{x}\right)-E(\beta_j(x_i))=B_1(j)h_j^2+o_p(h_j^2). $$
\noindent
For $h_j \propto n^r$ we have thus 
\begin{eqnarray*}
\sqrt n& E&\left(\frac{1}{n}\sum_{i=1}^n\hat{\beta}_j^{h_j}(x_i)\Big|\mathbf{x}\right)-\sqrt n E(\beta_j(x_i))\\
&=&n^{1/2}B_1(1)n^{2r}+ o_p(n^{1/2}n^{2r}) \\
&=&O(n^{1/2+2r})+ o_p(n^{1/2}n^{2r}).
\end{eqnarray*}
Furthermore from (\ref{condvar.eq}) and for $h_j \propto n^r$  we can write
\begin{eqnarray*}
&&nVar\left(\frac{1}{n}\sum_{i=1}^n\hat{\beta}_j^{h_j}(x_i)\Big|\mathbf{x}\right)\\
&=&V_1(j)+\frac{1}{n^{1+r}}V_2(j)+n^{2r}V_3(j)+
o_p(1+n^{-1-r}+n^2) \\
&=&V_1(j)+O(n^{-1-r})+O(n^{2r})+o_p(1+n^{-1-r}+n^2).
\end{eqnarray*}

\bibliographystyle{chicago}
\bibliography{JHXdLarXiv}

\end{document}